\documentclass[pra,12pt, showpacs,onecolumn,floatfix, groupaddress,superscriptaddress]{revtex4}
\usepackage{times,amsmath,amsfonts,amssymb,latexsym}
\usepackage{graphicx,epsf}
\usepackage{epstopdf}
\usepackage{subfigure}
\newcommand{\ra}{{\rightarrow}}
\newcommand{\be}{\begin{equation}}
\newcommand{\ee}{\end{equation}}
\newcommand{\ba}{\begin{eqnarray}}
\newcommand{\ea}{\end{eqnarray}}
\newcommand\tr{{\mbox{Tr\,}}}
\newcommand{\ignore}[1]{}

\newcommand{\ket}[1]{\left | {#1} \right \rangle }

\newcommand{\bra}[1]{\left \langle {#1} \right | }

\newcommand{\st}{S_{top}}
\newcommand{\range}[2]{\{#1,\ldots,#2\}}
\newcommand{\ud}{\textrm{d}}
\newcommand{\im}{\textrm{im}}

\def\CC{{\rm\kern.24em \vrule width.04em height1.46ex depth-.07ex
    \kern-.30em C}}
\def\P{{\rm I\kern-.25em P}}
\def\RR{{\rm
         \vrule width.04em height1.58ex depth-.0ex
         \kern-.04em R}}

\def\bbbc{{\mathchoice {\setbox0=\hbox{$\displaystyle\rm C$}\hbox{\hbox
to0pt{\kern0.4\wd0\vrule height0.9\ht0\hss}\box0}}
{\setbox0=\hbox{$\textstyle\rm C$}\hbox{\hbox
to0pt{\kern0.4\wd0\vrule height0.9\ht0\hss}\box0}}
{\setbox0=\hbox{$\scriptstyle\rm C$}\hbox{\hbox
to0pt{\kern0.4\wd0\vrule height0.9\ht0\hss}\box0}}
{\setbox0=\hbox{$\scriptscriptstyle\rm C$}\hbox{\hbox
to0pt{\kern0.4\wd0\vrule height0.9\ht0\hss}\box0}}}}
\def\bbbz{{\mathchoice {\hbox{$\sf\textstyle Z\kern-0.4em Z$}}
{\hbox{$\sf\textstyle Z\kern-0.4em Z$}}
{\hbox{$\sf\scriptstyle Z\kern-0.3em Z$}}
{\hbox{$\sf\scriptscriptstyle Z\kern-0.2em Z$}}}}

\begin{document}

\title{Topological order, entanglement, and quantum memory at finite temperature}
\author{Dalimil Maz\'{a}\v{c}}
\affiliation{Perimeter
Institute for Theoretical Physics, 31 Caroline St. N, N2L 2Y5,
Waterloo ON, Canada}
\author{Alioscia Hamma}

\affiliation{Perimeter
Institute for Theoretical Physics, 31 Caroline St. N, N2L 2Y5,
Waterloo ON, Canada}

\begin{abstract}
We compute the topological entropy of the toric code models in arbitrary dimension at finite temperature. We find that the critical temperatures for the existence of full quantum (classical) topological  entropy correspond to the confinement-deconfinement transitions in the corresponding $\mathbb{Z}_{2}$ gauge theories. This implies that the thermal stability of topological entropy corresponds to the stability of quantum (classical) memory. The implications for the understanding of ergodicity breaking in topological phases are discussed. 
\end{abstract}

\maketitle
\section{Introduction}\label{sec:introduction} 
Topologically ordered (TO) states are ground states of certain quantum many-body systems that exhibit an order which does not rely on symmetry-breaking mechanism, and thus cannot be characterized by a non vanishing local order parameter \cite{wenbook}. They possess a ground state degeneracy which depends on the topology of the underlying space and which can not be lifted by local perturbations of the Hamiltonian, and a pattern of long-range entanglement.

For  pure states, we say they possess topological order if they span a degenerate ground space with a gap, and such that distinct ground states are locally indistinguishable, i.e. the reduced density matrix over any topologically trivial region does not depend on the choice of the state within the ground space \cite{hamma_05, bravyi, symm}. This property implies a topological robustness under local perturbations, which has made this kind of order interesting for  quantum computation \cite{Kitaev, tcreview}. This robustness property means that topological order is a property of a whole phase, and one is interested in some quantity that can label and detect this kind of order. It turns out that TO states are characterized by specific entanglement properties. First, they posses an area law with a finite universal correction \cite{hiz1,hiz2, Kitaev2006a, Levin2006}. This correction, called {\em Topological  Entropy} (TE), has been shown to be constant within the whole TO phase of the toric code \cite{hamma_08}, and therefore can be used as an order parameter to label the TO phases.  It has been shown that this quantity is also characteristic of other TO phases like the Kitaev honeycomb model \cite{qi, honey}, topological dimer phases,  fractional quantum Hall liquids \cite{Haque2007, dimer}, and TO phases with finite correlation length \cite{tefinite}, or other quantum states that are defined in group theoretic terms \cite{hiz3}. Also, these properties show up in the entanglement spectrum. To what extent which TO phases can be classified using the full entanglement spectrum is still an open problem \cite{renyi, haas, Li2008a}. Moreover, quantum phases break in domains of quasi adiabatic continuity, namely the set of those states that can be connected by means of evolution with a local Hamiltonian without closing a gap \cite{quasiadiabatic}. If we ignore symmetry, all the non TO states are in the same phase as the completely factorized state. In this sense, non TO states have trivial entanglement, while TO states belong to different classes of non trivial, long-range entanglement \cite{chen_gu_wen_10}.

However, a real physical system is always found at some finite temperature $T$ by coupling to a thermal environment and so does not occupy solely the ground state. If topological order is to be a physical phenomenon, it must therefore exist in thermal states at nonzero temperature too. We should then seek generalizations of the above characterizations of topological order to $T>0$. The given definition, in terms of mass gap and ground-state degeneracy, fails to generalize straightforwardly as we can not speak about locally indistinguishable distinct states at $T>0$ simply because the thermal state is unique. Nevertheless, we can generalize this definition through its physical implications. At zero temperature, it implies that a TO system can support long-lived quantum memory. Indeed, if we add a local perturbation to a Hamiltonian of such system of linear size $L$, we must go to $O(L)$-th order in perturbation theory to connect orthogonal ground states. Hence, the tunnelling amplitude between distinct ground states is $O(\exp(-L))$, in which case we obtain a quantum memory register with lifetime $\tau=O(\exp(L))$, which we call {\em stable} quantum memory. Similarly, a stable classical memory is a system in which we can reliably encode classical information for exponentially long times. We can generalize this  viewpoint to finite temperature by defining a topologically ordered system to be one which supports quantum memory with lifetime that scales exponentially with the size of the system. As an example, the toric code in 2D \cite{Kitaev} does not support any kind of memory, quantum or classical at any finite temperature $T>0$, while the Ising model in 2D and the toric code in 3D both have a critical temperature $T_c$ below which classical memory is stable \cite{bt, yoshida, castelnovo08-3d}.

The notion of TE generalizes to $T>0$ immediately\cite{castelnovo08-3d}. Scaling of the von Neumann entropy with subsystem proportions is more complicated then the area law at zero temperature, but one may take a suitable linear combination of von Neumann entropies of different subsystems and isolate the universal constant piece coming from structured entanglement \cite{Kitaev2006a,Levin2006}. For the toric code in 2D and 3D, the calculation of TE at finite temperature was performed in \cite{castelnovo07-2d} and \cite{castelnovo08-3d}. In 2D there is no TE at any finite temperature, just like there is no stable information. In 3D, there is a stable TE of completely classical origin, just like there is a stable classical memory. This fact strongly pushes the question whether the stability of TE and memory at finite temperature are always related, and if yes, why. 

Moreover, the characterization of TO as states with non trivial entanglement (NT) has recently been generalized to finite temperature by Hastings \cite{hastings_11}, by considering equilibrium states that cannot be connected by means of a quantum circuit of finite range to a mixed state which is made of product states in the energy eigenbasis. Non TO states at finite temperature do possess trivial entanglement (FAC). 

In this paper, we investigate the important question whether these three characterizations of topological order are still equivalent at $T>0$. We study the simplest model with TO, the Toric Code -in arbitrary $D$ spatial dimensions- introduced by Kitaev \cite{Kitaev}, which in the low energy sector realizes the $\mathbb{Z}_{2}$ lattice gauge theory.    We analyze the stability of quantum memory,  calculate the TE and compare these results also with the presence of NT or FAC as indicators of topological order. The calculation of TE at arbitrary temperature in the thermodynamic limit is made possible by decomposition into contributions from the two kinds of defects \cite{castelnovo08-3d} and a mapping to the $\mathbb{Z}_{2}$ lattice gauge theory. We find the critical temperatures for the stability of TE, corresponding to the confinement-deconfinement transitions of the underlying gauge theory \cite{fradkin}.

We find that the value of TE contains all the information about stability of quantum and classical memory in these models (and about the triviality of entanglement) and hence that all define the same notion of topological order. So, at least in the toric code models, we see that {\em quantum TE means stable quantum memory, while classical TE means stable classical memory}. We are also able to elucidate that the same physical mechanism is responsible for destruction of quantum memory and quantum TE.

The toric codes examined in this paper depend on two couplings $\lambda$ and $\mu$.  We find there are two critical temperatures $T_\lambda$ and $T_\mu$, such that the stability of memory and TE are connected as follows.
\begin{itemize}
 \item quantum memory stable and $\st(T)=\st(0)$ for $T\in[0,\min(T_\lambda,T_\mu))$, and NT
 \item classical memory stable and $\st(T)=\st(0)/2$ for $T\in(\min(T_\lambda,T_\mu),\max(T_\lambda,T_\mu))$, and FAC
 \item no stable memory and $\st(T)=0$ for $T\in(\max(T_\lambda,T_\mu),\infty)$, and FAC,
\end{itemize}
where $\st(T)$ is the topological entropy at temperature $T$. 

As usual in statistical mechanics, the existence of finite critical temperature depends on the dimensionality of the system, where low dimensional systems are less likely to have finite-temperature phase transitions. As particular cases, we recover the results of Castelnovo and Chamon \cite{castelnovo07-2d,castelnovo08-3d} in two and three dimensions. This precise correspondence leads us to conjecture that it holds in general TO systems and hence that we may define topological order at $T>0$ as follows: A thermal state is TO at $T>0$ if $\st(T)=\st(0)>0$.

The paper is organized as follows. In Section \ref{sec:model}, the general toric codes are described, together with their dualities and connections to lattice gauge theory. Statistics of defects is then used to analyse the stability of quantum and classical memory in the toric codes in Section \ref{sec:memory}. Decomposition of the topological entropy, duality of the toric codes and a map to $\mathbb{Z}_{2}$ lattice gauge theory lead to the calculation of TE in Section \ref{sec:entropy}. Our results are discussed and compared to Hastings's circuit definition in Section \ref{sec:discussion}, together with suggestions for further work.

\section{General toric codes}\label{sec:model}
\subsection*{The models}
The toric codes considered in this paper are labelled by a pair $(D,k)$, where $D$ is the dimension of the lattice and $k\in\range{1}{D-1}$. The toric code labelled by $(D,k)$ will be denoted $\mathcal{T}^{(D,k)}$ and is defined as follows. Let $\Lambda$ be a $D$-dimensional cubic lattice of linear size $L$ with periodic boundary conditions and let us refer to its elementary $k$-dimensional blocks as $k$-cells. Let us denote $P_{k}(\Lambda)$ the set of $k$-cells and $N=L^D$ the total number of 0-cells in our lattice, so that $|P_{k}(\Lambda)| = \binom{D}{k}N$. To obtain $\mathcal{T}^{(D,k)}$, put a spin-$1/2$ degree of freedom on each $k$-cell and associate the star operator $A_a = \otimes_{i|a\in\partial i}X_i$ with each $(k-1)$-cell $a$, where the product runs over all $k$-cells neigbouring $a$, and plaquette operator $B_b = \otimes_{i\in\partial b}Z_i$ with each $(k+1)$-cell $b$, where the product runs over the $k$-cells contained in $b$. $X_i, Z_i$ are the local Pauli spin operators. The Hamiltonian of $\mathcal{T}^{(D,k)}$ is
\begin{equation} \label{eqn:hamiltonian0} 
 H^{(D,k)}(\lambda,\mu) = -\lambda \sum_{a\in P_{k-1}(\Lambda)} A_a - \mu \sum_{b\in P_{k+1}(\Lambda)} B_b\,,
\end{equation}
where $\lambda,\mu>0$.

$A_a$ and $B_b$ overlap only if $a$ is contained in $b$, but then they share precisely $2$ $k$-cells, so that $[A_a,B_{b}] = [A_a,A_{a'}] = [B_b,B_{b'}] = 0$. Hence the ground state subspace is $\mathcal{H}_g = \{\ket{\psi}: A_a\ket{\psi} = B_b\ket{\psi} = \ket{\psi}\;\forall a,b\}$. Denote the dual lattice by $\Lambda^*$. The algebra $\mathcal{A}_c$ of operators commuting with the Hamiltonian is generated by products of $X$ over closed $(D-k)$-chains in $\Lambda^*$ and products of $Z$ over closed $k$-chains in $\Lambda$. The algebra $\mathcal{A}_t = \{\mathcal{O}:\mathcal{O}\ket{\psi} = \ket{\psi}\;\forall\ket{\psi}\in\mathcal{H}_g\}$ is generated by $A$ and $B$ operators and hence consists of products of $X$ over boundary $(D-k)$-chains in $\Lambda^*$ and products of $Z$ over boundary $k$-chains in $\Lambda$. Hence the algebra of operators acting on the ground state $\mathcal{A}_g=\mathcal{A}_c/\mathcal{A}_t$ is generated by products of $X$ over $(D-k)$-homologies of $\Lambda^*$ and products of $Z$ over $k$-homologies of $\Lambda$. The $k$th $\mathbb{Z}_{2}$ homology group of the $D$-dimensional torus is $H_k(T^D,\mathbb{Z}_{2}) = (\mathbb{Z}_{2})^{\binom{D}{k}}$. It is not hard to see that there is a canonical one-to-one correspondence between the $(D-k)$-homologies of $\Lambda^*$ and $k$-homologies of $\Lambda$, so that if we define the logical operators $\bar{X}_\alpha, \bar{Z}_\alpha$ for $\alpha=1,\ldots,\binom{D}{k}$ to be the described products over homology representatives, we can choose the $\alpha$ labels so that
\begin{equation} \label{eq:gsalgebra} 
  \bar{X}_\alpha\bar{Z}_\alpha=-\bar{Z}_\alpha\bar{X}_\alpha,\quad
  [\bar{X}_\alpha,\bar{Z}_\beta] = 0\quad\textrm{if }\alpha\neq\beta
\end{equation}
The algebra of Eq.\eqref{eq:gsalgebra} is just the algebra of $\binom{D}{k}$ independent spin-$1/2$ particles, so we find $\dim\mathcal{H}_g = 2^{\binom{D}{k}}$.

The information stored in the ground space is topologically protected since only products of $X$ and $Z$ over surfaces with nontrivial homology act nonidentically on $\mathcal{H}_g$. Considering a local perturbation at $T=0$, we would have to go to the $O(L)$-th order in perturbation theory to get a nonvanishing matrix element between orthogonal ground states, i.e. the tunnelling amplitude is exponentially small in the size of the system. The situation is very different at $T>0$ where stability of TO depends on the energy barrier for these defects to wind around the torus.

\subsection*{Duality}
The $\mathcal{T}^{(D,k)}$ model is exactly dual to the $\mathcal{T}^{(D,D-k)}$ model on the dual lattice, which will prove very useful in the following. To obtain this result, note first that each $j$-cell in $\Lambda$ intersects precisely one $(D-j)$-cell of $\Lambda^*$, so if $a\in P_j(\Lambda)$, let $a^*$ be the corresponding $(D-j)$-cell in $\Lambda^*$. Hence the spins naturally live on the $(D-k)$-cells of $\Lambda^*$. We now observe that if $e\in P_{j}(\Lambda)$ and $f\in P_{j+1}(\Lambda)$, then
\begin{equation}
 e\in\partial f\quad\Leftrightarrow\quad f^*\in\partial e^*,
\end{equation}
where $\partial$ denotes the boundary operator. Define $U$ to be the unitary operator swapping globally the $x$ and $z$ computational bases. Denoting $A^*_c$, $B^*_d$ the analogous $A$, $B$ operators on the dual lattice, where $c\in P_{D-k-1}(\Lambda^*)$ and $d\in P_{D-k+1}(\Lambda^*)$, we find
\begin{equation}
 UA_{a}U^{\dagger} = B^*_{a^*},\quad UB_{b}U^{\dagger} = A^*_{b^*}.
\end{equation}
Consequently, if
\begin{equation}
H_{*}^{(D,D-k)}(\lambda,\mu) = -\lambda \sum_{c\in P_{D-k-1}(\Lambda^*)} A^*_c - \mu \sum_{d\in P_{D-k+1}(\Lambda^*)} B^*_d
\end{equation}
is the Hamiltonian of the $\mathcal{T}^{(D,D-k)}$ on the dual lattice, the duality is expressed through the equation
\begin{equation}
 UH^{(D,k)}(\lambda,\mu)U^{\dagger} = H^{(D,D-k)}_{*}(\mu,\lambda).
\end{equation}
It follows that any thermal expectation values calculated in the $\mathcal{T}^{(D,k)}$ and $\mathcal{T}^{(D,D-k)}$ at the same temperature are connected by swapping $\lambda$ and $\mu$ since the two density matrices are conjugate.

The unique toric code in two dimension is the well-known $\mathcal{T}^{(2,1)}$ model with star and plaquette operators, which is self-dual. It also follows that there is only one kind of a toric code in $3$D, since $\mathcal{T}^{(3,1)}$ is dual to $\mathcal{T}^{(3,2)}$, and so we need to go to at least $4$D to find distinct models with equal $D$. It will become clear that the most interesting $4$D toric code is the self-dual $\mathcal{T}^{(4,2)}$.

\subsection*{$\mathbb{Z}_{2}$ lattice gauge theory}
In the limit $\lambda\rightarrow\infty$, the condition
\begin{equation}\label{eq:gauge_constraint} 
A_a\ket{\psi}=\ket{\psi}\;\forall a\in P_{k-1}
\end{equation}
is enforced on all physical states $\ket{\psi}$ projecting  $\mathcal{T}^{(D,k)}$ onto the corresponding $\mathbb{Z}_{2}$ pure lattice gauge theory \cite{wegner_71,kogut_79, fradkin} with the Hamiltonian
\begin{equation}
 H^{(D,k)}_{g} = - \mu \sum_{b\in P_{k+1}(\Lambda)} B_b\,,
\end{equation}
and gauge constraints \eqref{eq:gauge_constraint}. Let us denote this theory by $\mathcal{GT}^{(D,k)}$. Spin flips cause excitations, which take the form of boundary $(D-k-1)$-chains in $\Lambda^*$. A boundary $(D-k-1)$-chain $C$ defines the $(k+1)$-chain $C^*$ of $(k+1)$-cells such that $B_b=-1$ for $b\in C^*$. Such $(k+1)$-cells will be called ``flipped cells'' and $C^*$ the ``flipped chain'' in the following.

Duality of $\mathcal{GT}^{(D,k)}$ and $\mathcal{GT}^{(D,D-k-2)}$ for $k\in\range{0}{D-2}$ can be used to show that $\mathcal{GT}^{(D,k)}$ for this range of $k$s is a two-phase system \cite{wegner_71}, where the Elitzur's theorem forbids existence of a local order parameter. In the low-temperature phase, defects, which are necessarily at least 1-dimensional for $k<D-1$, are confined, and become deconfined at a finite critical temperature. $\mathcal{GT}^{(D,D-1)}$ contains $0$-dimensional defects, which become deconfined already at $T=0$, and thus the system only has the disordered phase.

\section{Quantum and classical memory in the toric codes}\label{sec:memory}
\subsection*{General discussion}
As noted in Section \ref{sec:model}, the primary interest in the toric codes stems from the capacity of the ground-state subspace to store qubits which are stable under local perturbations at zero temperature\cite{dklp_02}. The problem of the stability of quantum memory at finite temperature is much more delicate and of great importance for both theoretical implications and practical reasons. In the view that the resilience of quantum memory at finite temperature is a property of the phase, one would expect that it is necessary to have a critical temperature below which memory is stable \cite{autocorrelations}. A remarkable fact of the memory encoded in the ground space of the toric codes is that their thermal stability can be studied through the confinement-deconfinement transition of the $\mathbb{Z}_{2}$ lattice gauge theory \cite{dklp_02, castelnovo08-3d}. In this discussion, it is important to distinguish between classical and quantum memory, which are defined as follows. A qubit is prepared in a superposition $\ket{\psi} = \sum_{i}c_i\ket{i}$, and coupling with a thermal bath is switched on. Let $\tau_{q}$ be the time scale at which the off-diagonal elements of the density matrix go to zero in some basis, i.e. the time at which the quantum correlations disappear. Similarly, let $\tau_{c}$ be the time scale when the diagonal elements change significantly, or equivalently when one loses even the classical probabilities. We say that our system posesses quantum memory if $\tau_q=O(\exp(L))$, where $L$ is the size of the system, and only classical memory if $\tau_q=O(1)$ and $\tau_c=O(\exp(L))$. Finally, the system has no memory if both $\tau_q,\tau_c=O(1)$ in the size of the system.

The ground state subspace of $\mathcal{T}^{(D,k)}$ is isomorphic to the Hilbert space of $\binom{D}{k}$ $2$-level systems, and its algebra of logical operators is generated by $\bar{X}_\alpha,\bar{Z}_\alpha$ for $\alpha\in\range{1}{\binom{D}{k}}$, which satisfy \eqref{eq:gsalgebra}. The different $\alpha$ sectors are equivalent and independent and we will restrict to $\alpha=1$ in the following and drop the indices. Let us choose a basis $\{\ket{0},\ket{1}\}$ for this tensor factor of $\mathcal{H}_g$ such that $\bar{Z}\ket{0}=\ket{0}$, $\bar{Z}\ket{1}=-\ket{1}$, $\bar{X}\ket{0}=\ket{1}$ and $\bar{X}\ket{1}=\ket{0}$, prepare qubit in a superposition $\ket{\psi} = c_0\ket{0} + c_1\ket{1}$ and turn on coupling with a thermal bath at temperature $T$. The coupling is assumed local and so $T>0$ will produce local defects of $A_a$ and $B_b$. Quantum memory is destroyed when the thermal defects can change the eigenvalue of either $\bar{X}$ or $\bar{Z}$, while classical memory is preserved when the eigenvalue of either $\bar{X}$ or $\bar{Z}$ is robust under the thermal defects \cite{alicki_07,alicki_09}. We have seen that $\bar{X}$ is a product of $X$ over a $(D-k)$-homology in $\Lambda^*$ and $\bar{Z}$ a product of $Z$ over the dual $k$-homology in $\Lambda$. Hence the eigenvalue $\bar{X}$ is fragile when the $B_b$ defects can wind around the torus. In turn, the deconfinement of these defects can be seen as the deconfined phase of the corresponding gauge theory, i.e. the deconfined phase of  $\mathcal{GT}^{(D,k)}$ with coupling $\mu$. Similarly, the eigenvalue of $\bar{X}$ is fragile when the $A_a$ defects can conspire to produce surfaces with nontrivial homology. The $\lambda\leftrightarrow\mu$ duality tells us this happens precisely when $\mathcal{GT}^{(D,D-k)}$ with coupling $\lambda$ is deconfined. We can now summarize these results in the following table
\begin{table}[h]
 \begin{tabular}{l c l}
  quantum memory &$\Leftrightarrow$& both $\mathcal{GT}^{(D,D-k)}(T/\lambda)$ and $\mathcal{GT}^{(D,k)}(T/\mu)$ confined\\
  only classical memory &$\Leftrightarrow$& either $\mathcal{GT}^{(D,D-k)}(T/\lambda)$ or $\mathcal{GT}^{(D,k)}(T/\mu)$ deconfined\\
  no memory &$\Leftrightarrow$& both $\mathcal{GT}^{(D,D-k)}(T/\lambda)$ and $\mathcal{GT}^{(D,k)}(T/\mu)$ deconfined
 \end{tabular}
\caption{Quantum and classical memory in $\mathcal{T}^{(D,k)}$}\label{tab:memory} \end{table}

\subsection*{Examples}
The lattice gauge theory 
$\mathcal{GT}^{(2,1)}(T/\mu)$ has phase transition at $T/\mu=0$, and so the 2D toric code has neither quantum nor classical memory at any finite temperature \cite{alicki_07}. $\mathcal{GT}^{(3,2)}(T/\lambda)$ also deconfines at $T/\lambda=0$, but $\mathcal{GT}^{(3,1)}(T/\mu)$ has a nontrivial low-temperature phase, so that the 3D toric code loses quantum memory at $T=0$, but preserves classical memory up to a finite critical temperature, which is proportional to the coupling of the plaquette operators. The simplest toric code with quantum memory at finite temperature is the $\mathcal{T}^{(4,2)}$ model \cite{dklp_02}, whose stability is controlled by $\mathcal{GT}^{(4,2)}(T/\mu)$ \cite{alicki_09}.

\section{Topological entropy}\label{sec:entropy}
\subsection*{Definition}
In this section, the topological entropy of $\mathcal{T}^{(D,k)}$ will be calculated for general couplings at any $T$ in the thermodynamic limit and shown to correspond precisely to the behaviour of quantum and classical memory as discussed in the previous section.

At $T=0$, topological entropy $\st(0)$ is defined as the universal part of bipartite entanglement entropy $S_{e}$, which does not scale with the subsystem boundary
\begin{equation}
 S_{e} = \alpha|\partial| - \st(0).
\end{equation}
As opposed to $\alpha$, it is robust under local perturbations \cite{hamma_08} and also $\st(0)\geq0$, so that it represents an order of entanglement.

\begin{figure}[htb] 
\includegraphics[width=\textwidth]{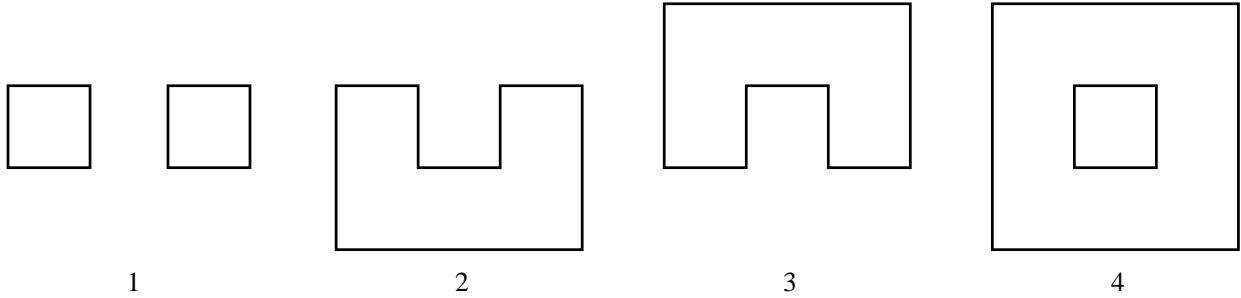}
\caption{Partitions for the $D=2$ model}\label{fig:2dpartitions}
\end{figure}

At $T>0$, the scaling of $S_{e}$ is more complicated and we need to extract the universal term $\st(T)$ by taking a linear combination \cite{Kitaev2006a,Levin2006}
\begin{equation}
 \st(T) = \sum_{i}\sigma(i)S_{e}(T,C_i),
\end{equation}
where $\sigma(i)$ are the signs of partitions $C_i\sqcup D_i = P_{k}(\Lambda)$, which are chosen so that the linear combination of the bulk and boundary chains of $C_i$ of any dimensionality are zero, and we are left with the topological contributions only. Here we generalize to arbitrary $D$ the clever construction of \cite{castelnovo08-3d}. For a general $\mathcal{T}^{(D,k)}$, we will define the $C_i^{(D)}$, $i\in\range{1}{4(D-1)}$ by induction on $D$ as follows. For $D=2$, we choose the four partition as in Figure \ref{fig:2dpartitions}, with $\sigma(2)=\sigma(3)=-\sigma(1)=-\sigma(4) = 1$, where the outer, inner squares in $C_4^{(2)}$ have sidelengths $a$, $a/3$ respectively. Having constructed all $C_i^{(D)}$, we define $C_{4D}^{(D+1)}$ to be the $(D+1)$-dimensional cube of side $a$ missing a $(D+1)$-dimensional cube of side $a/3$ from its middle. Analogously to Fig.\ref{fig:2dpartitions}, we define $C_{4D-1}^{(D+1)}$, $C_{4D-2}^{(D+1)}$ as the upper and lower two-thirds of $C_{4D}^{(D+1)}$, upper and lower meant in the last dimension, and $C_{4D-3}^{(D+1)} = C_{4D-1}^{(D+1)}\cap C_{4D-2}^{(D+1)}$. For $i\in\range{1}{4(D-1)}$, construct $C_i^{(D+1)} = C_i^{(D)}\times I_{a/3}$, where the interval $I_{a/3}$ of length $a/3$ is added in the last dimension so that $C_{4D-4}^{(D+1)}=C_{4D-3}^{(D+1)}$. Finally, choose $\sigma(i)$
\begin{equation}
 \sigma(i) = \begin{cases}
	      -1\quad\textrm{if }i\equiv 0,1\!\mod 4\\
              +1\quad\textrm{if }i\equiv 2,3\!\mod 4
	      \end{cases}.
\end{equation}

\begin{figure}[htb] 
\includegraphics[width=\textwidth]{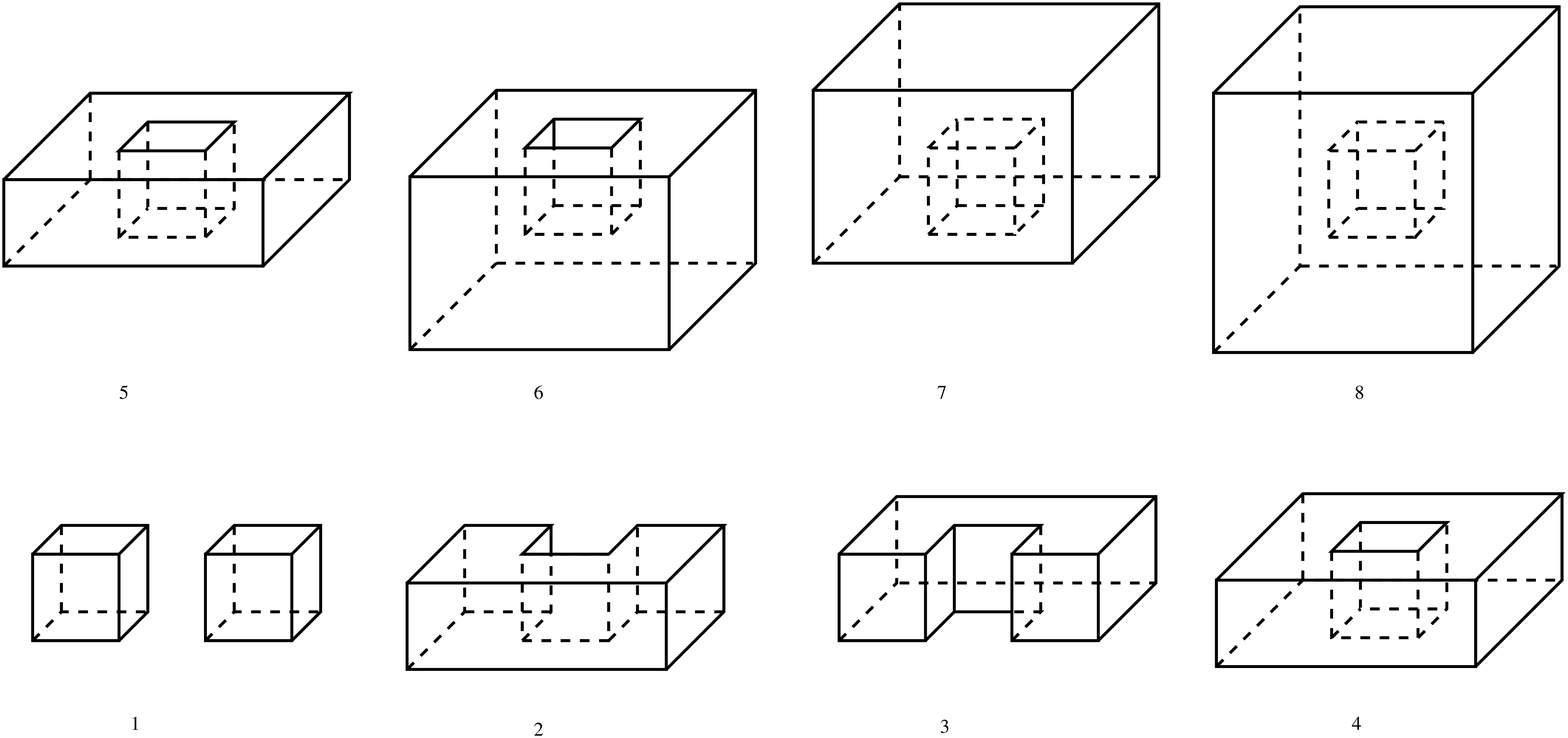}
\caption{Partitions for the $D=3$ model}\label{fig:3dpartitions}
\end{figure}

\begin{figure}[htb] 
\includegraphics[width=\textwidth]{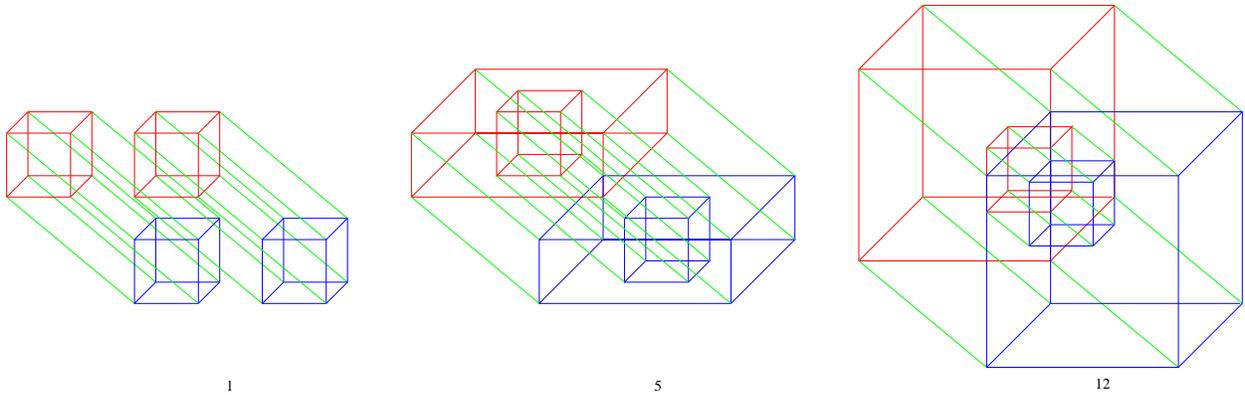}
\caption{Partitions $i=1,5,12$ for the $D=4$ model}\label{fig:4dpartitions}
\end{figure}

As an example, for $D=3$ and $D=4$ we have the partitions of Fig.\ref{fig:3dpartitions} and \ref{fig:4dpartitions}. The fourth dimension is represented by the green segments in Fig. \ref{fig:4dpartitions}, while the blue and red are 3-boundaries living in the first three dimensions. This choice of partitions was motivated by the requirement that the signed bulk and boundary chains add up to zero. Indeed, they clearly do for $D=2$, and hence by induction on $D$, 
\begin{equation}
\sum_{i=1}^{4(D-1)}\sigma(i)C^{(D+1)}_i = 0,
\end{equation}
since $C_i^{(D+1)} = C_i^{(D)}\times I_{a/3}$ for $i\in\range{1}{4(D-1)}$. Moreover, it follows straight from the definition that
\begin{equation}
 \sum_{i=4D-3}^{4D}\sigma(i)C_i^{(D+1)} = 0,
\end{equation}
and hence the full set of partitions gives zero net bulk chain
\begin{equation}
 \sum_{i=1}^{4D}\sigma(i)C_i^{(D+1)} = 0.
\end{equation}  
The same argument also works for the non oriented boundaries with signs $\sigma(i)$. When choosing the partitions, we also required that their collection is symmetric under the exchange $C_i\leftrightarrow D_i$, besides the global torus topology. For example, $C^{(D)}_1$ has $2$ connected components and so in $D$ dimensions, we are forced to introduce the $i=4(D-1)$ partition, where $D^{(D)}_{4(D-1)}$ also has two connected components. Finally, if we also demand that for each $d\in\range{0}{D-1}$, there is an $i$ such that $C_i$ has nontrivial homology of dimension $d$, our choice is a very natural one. In the thermodynamic limit, we scale both $L,a\rightarrow\infty$.
\subsection*{Zero temperature}
Let us first calculate $\st$ at zero temperature as this result forms the core of the calculation at $T>0$. Define the groups of spin flips $G=\langle A_a|\,a\in P_{k-1}\rangle$, $G_i = \{g\in G|\, g_{D_i} = 1_{D_i}\}$ and $H_i = \{g\in G|\, g_{C_i} = 1_{C_i}\}$, where $g_{D_i}$ denotes restriction of $g$ to $D_i$ and $1_{D_i}$ is the identity transformation on $D_i$. Since any ground state is a uniform superposition over the group $G$, the entanglement entropy of partition $i$ is \cite{hamma_05}
\begin{equation}
 S^{(D,k)}_i = \log\left(\frac{|G|}{|G_i||H_i|}\right).
\end{equation}
Hence the TE is
\begin{equation}\label{eq:st01} 
 \st^{(D,k)}(0) = -\sum_{i=1}^{4(D-1)}\sigma(i)\log\left(|G_i||H_i|\right),
\end{equation}
because there is an equal number of positive and negative partitions. Now we turn to the calculation of $|G_i|$ and $|H_i|$. The contribution from the global topology of $\Lambda$ to $\log(|H_i|)$ is the same for all $i$ and therefore cancels in \eqref{eq:st01}. It will therefore be omitted in some of the following equalities, which we will write as $\doteq$ for that reason. $G_i$ consists of those closed $(D-k)$-chains in $C_i^{*}$, which are also boundaries of $(D-k+1)$-chains in $\Lambda^*$. The elements of $G_i$ which are not boundaries of chains in $C_i$ must arise from nontrivial $(k-1)$-homologies of $D_i^*$, so that 
\begin{equation}
 \log(|G_i|) \doteq \log(|H_{k-1}(D_i^{*})|) + \log(|B_{D-k}(C_i^{*})|),
\end{equation}
where $H_{d}$, $B_{d}$ are the $\mathbb{Z}_{2}$ homology, boundary groups respectively. Let $X_d$ and $Z_d$ be respectively the groups of all, and closed $\mathbb{Z}_{2}$ $d$-chains in $C_i$. In other words, $Z_d = \ker(\partial_d)$, $B_{d} = \im(\partial_{d+1})$, where $\partial_d: X_d\ra X_{d-1}$ is the boundary homomorphism. Then the $\mathbb{Z}_{2}$ homology groups are defined by
\begin{equation}
 H_d = Z_d/B_d.
\end{equation} 
By the first isomorphism theorem applied iteratively to $\partial_d$ for $d\in\range{D-k+1}{D}$, we obtain
\begin{equation}
 \log_2(|G_i|) \doteq b_{k-1}(D_i^{*}) + \sum_{j=1}^{k}(-1)^j[b_{D-(k-j)}(C_i^{*})- |P_{D-(k-j)}(C_i^{*})|],
\end{equation}
where $b_{j}$ is the $j$-th Betti number. Similar analysis holds for $|H_i|$ and we arrive at the result
\begin{align} \label{eq:st02} 
 \frac{\st(0)}{\log 2} &= -\sum_{i}\sigma(i)\left\{b_{k-1}(D_i) + b_{k-1}(C_i) +\vphantom{\sum_{j=1}^{k}}\right.\nonumber\\
 &\left.+\sum_{j=1}^{k}(-1)^j[b_{D-(k-j)}(C_i) + b_{D-(k-j)}(D_i)]\right\},
\end{align}
where the terms containing the $j$-cell numbers $|P_j|$ subtracted since the bulks of signed partitions add up to zero, and the stars were dropped since $C_i$, $D_i$ have the same topology as $C_i^{*}$, $D_i^{*}$. Our task is reduced to calculating the Betti numbers of $C_i$ and $D_i$. Define reduced Betti numbers $b'_0 = b_0-1$ and $b'_i = b_i$ for $i>0$. They are
\begin{align}
b'_i(C_{4j}) &= \begin{cases}
                 1\quad\textrm{if }i=j\\
		 0\quad\textrm{otherwise}
                \end{cases}\\
b'_i(C_{4j+1}) &= \begin{cases}
                 1\quad\textrm{if }i=j\\
		 0\quad\textrm{otherwise}
                \end{cases}\\
b'_i(C_{4j+2}) &= b'_i(C_{4j+3}) = 0\\
b'_i(D_{4j}) &\doteq \begin{cases}
                 1\quad\textrm{if }i=D-j-1\\
		 0\quad\textrm{otherwise}
                \end{cases}\label{eq:bettiD1} \\
b'_i(D_{4j+1}) &\doteq \begin{cases}
                 1\quad\textrm{if }i=D-j-1\\
		 0\quad\textrm{otherwise}
                \end{cases}\label{eq:bettiD2}\\
b'_i(D_{4j+2}) &= b'_i(D_{4j+3}) \doteq 0.\label{eq:bettiD3}
\end{align}
Substituting these values into \eqref{eq:st02}, we find that
\begin{equation}
 \st^{(D,k)}(0) = 2\log 2,
\end{equation}
independently of $D$ and $k$, which also reproduces the known results for 2D and 3D \cite{castelnovo07-2d,castelnovo08-3d}. Had we separated the contribution from $|G_i|$ and $|H_i|$, we would have found
\begin{equation}\label{eq:gisum} 
 -\sum_{i=1}^{4(D-1)}\sigma(i)\log\left(|G_i|\right) = -\sum_{i=1}^{4(D-1)}\sigma(i)\log\left(|H_i|\right) = \log 2.
\end{equation}

\subsection*{Finite Temperature}
Let us now proceed by calculating $\st^{(D,k)}(T,\lambda,\mu)$ for general $T$. To perform this calculation, we will use the key property found in  \cite{castelnovo08-3d}, where it was shown that the entanglement entropy in general toric codes decomposes into a sum of two terms, with the first coming from the star operators and depending only on $T/\lambda$ and the second from plaquette operators depending on $T/\mu$. Thus we may write
\begin{equation}
 S^{(D,k)}_{e}(T,\lambda,\mu) = Q_{e}^{(D,D-k)}(T/\lambda) + R_{e}^{(D,k)}(T/\mu),
\end{equation}
where the reason for notation $Q_e^{(D,j)}$, $R_e^{(D,j)}$ will become clear shortly. TE is linear in von Neumann entropies, so the same factorization applies
\begin{equation}
 \st^{(D,k)}(T,\lambda,\mu) = Q_{top}^{(D,D-k)}(T/\lambda) + R_{top}^{(D,k)}(T/\mu).
\end{equation}
We have seen in Section \ref{sec:model} that $\mathcal{T}^{(D,k)}$ with couplings $(\lambda,\mu)$ is equivalent to the $\mathcal{T}^{(D,D-k)}$ with couplings $(\mu,\lambda)$ on the dual lattice. Since $C_i$, $D_i$ have the same topology as $C_i^{*}$, $D_i^{*}$, the same duality holds for TE
\begin{equation}
 \st^{(D,k)}(T,\lambda,\mu) = \st^{(D,D-k)}(T,\mu,\lambda).
\end{equation}
It follows that $Q_{top}^{(D,j)}$ and $R_{top}^{(D,j)}$ can be chosen to coincide
\begin{equation}
 Q_{top}^{(D,j)}(x) = R_{top}^{(D,j)}(x)
\end{equation}
and we can write
\begin{equation}\label{eq:stT1} 
 \st^{(D,k)}(T,\lambda,\mu) = Q_{top}^{(D,D-k)}(T/\lambda) + Q_{top}^{(D,k)}(T/\mu).
\end{equation}
We have thus reduced our problem to finding the functions $Q_{top}^{(D,j)}(x)$ for $j\in\range{1}{D-1}$. The zero-temperature result tells us that we must have $Q_{top}^{(D,j)}(0) = \log 2$. The $\lambda\rightarrow\infty$ limit of equation \eqref{eq:stT1} then yields
\begin{equation} \label{eq:qtop0} 
 Q_{top}^{(D,k)}(T/\mu) = \lim_{\lambda\rightarrow\infty}\st^{(D,k)}(T,\lambda,\mu) - \log 2.
\end{equation}
But $\lim_{\lambda\rightarrow\infty}\st^{(D,k)}(T,\lambda,\mu)$ is nothing but the topological entropy in the gauge theory $\mathcal{GT}^{(D,k)}(T/\mu)$ since $\lambda\rightarrow\infty$ imposes the gauge constraints $A_a = 1$.

\subsection*{Topological entropy in the gauge theory}
In order to compute $S_{gtop}^{(D,k)}(T/\mu)\equiv\lim_{\lambda\rightarrow\infty}\st^{(D,k)}(T,\lambda,\mu)$, introduce the projection operator $P$ onto the states of the gauge theory
\begin{equation}
 P = \frac{1}{|G|}\sum_{g\in G}g.
\end{equation}
The thermal density matrix of our model is
\begin{equation}
 \rho(T/\mu) = \frac{1}{Z}\exp\left(-\beta\mu\sum_{b\in P_{k+1}(\Lambda)}B_b\right)P,
\end{equation}
where (in the computational $z$-basis)
\begin{align}
 Z &= \tr\left[\exp\left(-\beta\mu\sum_{b\in P_{k+1}(\Lambda)}B_b\right)P\right]= \nonumber\\
 &= \frac{1}{|G|}\sum_{\substack{g\in G\\f\in F}}\bra{0}fgf\ket{0}e^{-\beta\mu M(f)} \nonumber\\
 &= \frac{1}{|G|}\sum_{f\in F}e^{-\beta\mu M(f)}
\end{align}
is the partition function of $\mathcal{GT}^{(D,k)}(T/\mu)$, where $F$ is the group of all spin flips on $P_{k}(\Lambda)$ and $M(f) = \bra{f}\sum_{b\in P_{k+1}(\Lambda)}B_b\ket{f}$ is the total plaquette magnetization of $f$. Let us also define the groups $F_i = \{f\in F|\, f_{D_i} = 1_{D_i}\}$ and $E_i = \{f\in F|\, f_{C_i} = 1_{C_i}\}$ that act nontrivially only on $C_i$, $D_i$ respectively. The reduced density matrix of subsystem $C_i$ is then
\begin{equation}\label{eq:rg} 
 \rho_{i}(T/\mu) = \frac{\sum_{\substack{g\in G_i\\f\in F}} e^{-\beta\mu M(f)}\left(g\ket{f}\bra{f}\right)|_{C_i}}{|G|Z},
\end{equation}
where $\mathcal{O}|_{C_i}$ denotes the projection of operator $\mathcal{O}$ onto the Hilbert space of $C_i$. Let us use the replica tric to find the entanglement entropy of $C_i$ in the gauge theory
\begin{equation}\label{eq:replica} 
 S_{gi}^{(D,k)}(T/\mu) = -\left.\frac{\ud}{\ud n}\right|_{n=1}\tr\left[\rho_{i}^{n}(T/\mu)\right].
\end{equation}
The trace of the $n$-th power of $\rho_{i}^n$ is found from \eqref{eq:rg} to be
\begin{equation}
 \tr\left[\rho_{C_i}^{n}(T/\mu)\right] = \frac{|G_{i}|^{n-1}}{|G|^nZ^n} \sum_{f_1,\ldots,f_n\in F} e^{-\beta\mu \sum_{m=1}^{n}M(f_m)}\delta_{C_i}(f_1,\ldots,f_n),
\end{equation}
where
\begin{equation}
 \delta_{C_i}(f_1,\ldots,f_n) = \begin{cases}
                                 1\quad\textrm{iff }f_1|_{C_i} = \ldots =f_n|_{C_i}\\
				 0\quad\textrm{otherwise}.
                                \end{cases}
\end{equation}
We can trivialize the delta constraint through substitution $f_m = fe_m$, where $f\in F_i$ and $e_m\in E_i$
\begin{align}\label{eq:trr} 
 \tr\left[\rho_{i}^{n}(T/\mu)\right] &= \frac{|G_{i}|^{n-1}}{|G|^nZ^n} \sum_{f\in F_i}\sum_{e_1,\ldots,e_n\in E_i} \exp\left[-\beta\mu \sum_{m=1}^{n}M(fe_m)\right] \nonumber\\
 &= \frac{|G_{i}|^{n-1}}{|G|^nZ^n} \sum_{f\in F_i}[q_i(f)]^n,
\end{align}
where
\begin{equation}
 q_i(f) = \sum_{e\in E_i}e^{-\beta\mu M(fe)}. 
\end{equation} 
Expression \eqref{eq:trr} is ready to be used in the replica tric \eqref{eq:replica}, with the result
\begin{equation}
 S_{gi}^{(D,k)}(T/\mu) = -\log(|G_i|) + \log(|G|Z) - \frac{1}{|G|Z}\sum_{f\in F_i}q_i(f)\log q_i(f).
\end{equation}
Observe that $q_i$ defines a probability distribution
\begin{equation}
 p_i(f)\equiv\frac{q_i(f)}{|G|Z}
\end{equation}
on the group $F_i$. Indeed, $\sum_{f\in F_i}q_i(f) = |G|Z$. Taking now the linear combination over the signed partitions $C_i\sqcup D_i$, the $i$-independent terms cancel and we arrive at
\begin{align}
 S_{gtop}^{(D,k)}(T/\mu) &= -\sum_{i=1}^{4(D-1)}\sigma(i)\log\left(|G_i|\right)  -\sum_{i=1}^{4(D-1)}\sigma(i)\left[\sum_{f\in F_i}p_i(f)\log p_i(f)\right] =\nonumber\\
 &=\log 2 - \sum_{i=1}^{4(D-1)}\sigma(i)\left[\sum_{f\in F_i}p_i(f)\log p_i(f)\right],
\end{align}
where \eqref{eq:gisum} was used in the second equality. The first term is just the contribution of $|G_i|$ to the zero temperature result \eqref{eq:st01}. Returning back to $Q_{top}^{(D,k)}(T/\mu)$, we find from equation \eqref{eq:qtop0}
\begin{equation}\label{eq:qtop} 
 Q_{top}^{(D,k)}(T/\mu) = - \sum_{i=1}^{4(D-1)}\sigma(i)\left[\sum_{f\in F_i}p_i(f)\log p_i(f)\right].
\end{equation} 

\subsection*{Isolating gauge redundancy}
Expression \eqref{eq:qtop} still includes some gauge redundancy. Define the groups of spin flips $\mathcal{G} = G\times\langle\bar{X}_{\alpha}|\alpha=1,\ldots,\binom{D}{k}\rangle$, $\mathcal{G}_i = \{g\in \mathcal{G}|\, g_{D_i} = 1_{D_i}\}$ and $\mathcal{H}_i = \{g\in \mathcal{G}|\, g_{C_i} = 1_{C_i}\}$ analogous to $G$, $G_i$ and $H_i$, but this time also containing the noncontractible flips. Magnetization $M(f)$ is invariant under $f\mapsto gf$ for $g\in\mathcal{G}$, so that for $f\in F_i$
\begin{equation}
 p_i(f) = \frac{|\mathcal{H}_{i}|}{|G|Z}\sum_{e\in \tilde{E}_i} e^{-\beta\mu M(fe)},
\end{equation}
where $\tilde{E}_i\equiv E_i/\mathcal{H}_i$. Moreover, if we define the group $K_i = \{f\in F_i|\,\exists e\in E_i : fe\in\mathcal{G}\}$, we find that $p_i(f)$ only depends on the coset $[f]\in \tilde{F}_i\equiv F_i/K_i$. Note that $|K_i| = |\mathcal{G}|/|\mathcal{H}_i|$, and hence
\begin{equation}
 \sum_{f\in F_i}p_i(f)\log p_i(f) = \log(|\mathcal{H}_{i}|) - \log(|\mathcal{G}|) + \sum_{f\in \tilde{F}_i}\tilde{p}_i(f)\log \tilde{p}_i(f),
\end{equation}
where $\tilde{p}_i(f)$ is the non-redundant probability distribution over $\tilde{F}_i$ given by
\begin{equation}
 \tilde{p}_i(f) \equiv \frac{1}{W} \sum_{e\in \tilde{E}_i} e^{-2\beta\mu \Phi(fe)},
\end{equation}
where $\Phi(f)$ is the number of flipped $(k+1)$-cells in configuration $f$ and
\begin{equation}
 W = \sum_{f\in \tilde{F}}e^{-2\beta\mu\Phi(f)},
\end{equation}
$\tilde{F}\equiv F/\mathcal{G}$. Elements of $\tilde{F}$ are precisely the physically distinct configurations of the gauge theory, and elements of $\tilde{F}_i$ are those configurations of $(k+1)$-cells inside $C_i$ which can be extended to consistent global configurations of the gauge theory, i.e. to boundary $(D-k-1)$-chains in $\Lambda^*$. Similarly, one should regard $\tilde{E}_i$ as containing distinct ways to complete the boundary $(D-k-1)$-chains inside $D_i^*$. We have now completely removed the gauge redundancy from our expressions.

Observe that by \eqref{eq:gisum}
\begin{equation}
 -\sum_{i=1}^{4(D-1)}\sigma(i)\log\left(|\mathcal{H}_i|\right) =
 -\sum_{i=1}^{4(D-1)}\sigma(i)\log\left(|H_i|\right) = \log2,
\end{equation}
and consequently
\begin{equation}\label{eq:qtop2} 
 Q_{top}^{(D,k)}(T/\mu) = \log2 -  \sum_{i=1}^{4(D-1)}\sigma(i)\left[\sum_{f\in \tilde{F}_i}\tilde{p}_i(f)\log \tilde{p}_i(f)\right].
\end{equation}
In other words, the gauge redundancy contributes the factor $\log2$ to TE. This factor can be identified with the $|H_i|$ part of \eqref{eq:st01}.

\subsection*{Phase transition}
At this point, we are finally ready to show the mechanism that produces a phase transition at a certain temperature $T=T_c$ for the topological entropy. To this end, we investigate the second term in Eq.\eqref{eq:qtop2} and show that it leads to a phase transition in $Q_{top}^{(D,k)}(T/\mu)$. First notice that at $T=0$, only $f=1\in\tilde{F}_i$ produces nonzero probability $\tilde{p}_i$, since all other configurations contain flipped $(k+1)$-cells. But then $\tilde{p}_i(1)=1$, and so indeed
\begin{equation}
 Q_{top}^{(D,k)}(0) = \log 2,
\end{equation}
as required by self-consistency. Let us rewrite the square bracket in \eqref{eq:qtop2} as
\begin{align}
 \sum_{f\in \tilde{F}_i}\tilde{p}_i(f)\log \tilde{p}_i(f) &= \sum_{\substack{f\in \tilde{F}_i\\e\in\tilde{E}_i}} \frac{1}{W}e^{-2\beta\mu \Phi(fe)}\log\left[\sum_{e'\in\tilde{E}_i}\frac{1}{W}e^{-2\beta\mu \Phi(fe')}\right] =\nonumber\\
 &= \sum_{f\in \tilde{F}} \frac{1}{W}e^{-2\beta\mu \Phi(f)}\log\left[\sum_{e''\in\tilde{E}_i}\frac{1}{W}e^{-2\beta\mu \Phi(fe'')}\right],
\end{align}
where we used $\tilde{F}=\tilde{F}_i\times\tilde{E}_i$ and the fact that multiplication by $e^{-1}\in\tilde{E}_i$ produces a mere permutation within $\tilde{E}_i$. All the dependence on $i$ is now inside the logarithm and thus
\begin{equation}\label{eq:qtop3} 
 Q_{top}^{(D,k)}(T/\mu) = \log 2 - \sum_{f\in\tilde{F}}r(f)\log\left[\frac{s_+(f)}{s_-(f)}\right],
\end{equation}
where
\begin{equation}
 r(f) = \frac{1}{W}e^{-2\beta\mu \Phi(f)}
\end{equation} 
and
\begin{equation}
 s_{\pm}(f) = \prod_{i|\sigma(i)=\pm1}\left[\sum_{e_i\in\tilde{E_i}}\exp(-2\mu\beta \Phi(fe_i))\right].
\end{equation}

Let us suppose that $\mathcal{GT}^{(D,k)}(T/\mu)$ is in the confined phase. Big membranes of defects are suppressed, i.e. $r(f)$ is exponentially small in the defect extent. In particular, defects can not detect the topology of $C_i$. When we expand $s_\pm(f)$ for fixed $f\in\tilde{F}$, we can see it is a sum over overlapping configurations of the gauge theory, differing from $f$ only in $D_i$s. For $f=1$, the allowed multiplicity of a defect at a given $b\in P_{k+1}(\Lambda)$ in a term of $s_\pm(f)$ is equal to $|\{i|\sigma(i)=\pm1\wedge b^*\in P_{D-k-1}(D_i^*)\}|$, for the plus and minus sign respectively. Since all contributing defects are local and the partitions satisfy the chain equation $\sum_{i|\sigma(i)=1}C_i = \sum_{j|\sigma(j)=-1}C_j$, we find that there is one-to-one correspondence between contributing terms in $s_{+}(f)$ and $s_{-}(f)$, and hence that
\begin{equation}
 s_{+}(f)=s_{-}(f)
\end{equation}
for any $f\in\tilde{F}$. Therefore, for $T/\mu<(T/\mu)_{crit}$
\begin{equation}
 Q_{top}^{(D,k)}(T/\mu) = \log 2.
\end{equation}

On the other hand, when $\mathcal{GT}^{(D,k)}(T/\mu)$ is deconfined, $T/\mu>(T/\mu)_{crit}$, a typical configuration will contain many topological defects, i.e.  $(D-k-1)$-branes (which are boundaries). By \eqref{eq:bettiD1}-\eqref{eq:bettiD3}, the only $D^*_i$s with nontrivial $(D-k-1)$-homology are those with $i = 4k$, $4k + 1$, where the later only exists if $k<D-1$ (the global toric topology does not count as nontrivial here as it contributes to all bipartitions in the same manner). Let $l\in\tilde{F}$ be a membrane which wraps around $C^*_{4k}$, i.e. a nontrivial $(D-k-1)$-homology in $D^*_{4k}$. Then $l\in\tilde{E}_i$ precisely for $i<4k$, hence
\begin{equation}
 |\{i|\sigma(i)=1\wedge l\in\tilde{E}_i\}| = |\{i|\sigma(i)=-1\wedge l\in\tilde{E}_i\}|+1.
\end{equation}
When $k=D-1$, every term in $s_{+}(f)$, after expanding the product over $i|\sigma(i)=1$, can be obtained by twice as many ways as the same term in $s_{-}(f)$. This is because $D^{*}_{4(D-1)}$ must contain an even number of topological defects (i.e. an even number of point-like defects inside the smaller hypercube), whereas any parity is allowed for $i<4(D-1)$. For general $k$, remember that above $T_c$ there is an infinite number of defects in the thermodynamic limit, and therefore on average again there are twice as many ways of distributing the topological defects among the positive partitions with respect to the negative ones.  Moreover, the variance of this distribution is zero in the thermodynamic limit \cite{footnote}. Hence, in the thermodynamic limit, above $T_c$, we have
\begin{equation}
 \frac{s_{+}(f)}{s_{-}(f)} = 2
\end{equation}
for any $f\in\tilde{F}$. To illustrate this point, consider the $\mathcal{T}^{(2,1)}$, where the defects are $0$-boundaries, i.e. pairs of points virtually connected by a line. Terms of $s_{-}(1)$ containing an odd number of endpoints inside the small square must all come from $i=1$, whereas in $s_{+}(1)$, they can come from both $i=2,3$. Similar reasoning applies when $f\neq 1$. Consequently, if $T/\mu>(T/\mu)_{crit}$, we obtain
\begin{equation}
 \sum_{f\in\tilde{F}}r(f)\log\left[\frac{s_+(f)}{s_-(f)}\right] = \log 2
\end{equation}
and the two terms in \eqref{eq:qtop3} precisely cancel. Hence
\begin{equation}
 Q_{top}^{(D,k)}(T/\mu) = \begin{cases}
                           \log2\quad&\textrm{if }\mathcal{GT}^{(D,k)}(T/\mu)\textrm{ confined}\\
			   0\quad&\textrm{if }\mathcal{GT}^{(D,k)}(T/\mu)\textrm{ deconfined},
                          \end{cases}
\end{equation}
i.e. $Q_{top}^{(D,k)}(T/\mu)$ experiences a phase transition at the same $T/\mu$ as the lattice gauge theory. We can finally substitute into \eqref{eq:stT1} to find the behaviour of TE as in Table \ref{tab:entropy}. The similarity with Table \ref{tab:memory} is striking.
\begin{table}[h] 
 \begin{tabular}{l c l}
  $\st^{(D,k)}(T,\lambda,\mu)=2\log2$ &$\Leftrightarrow$& both $\mathcal{GT}^{(D,D-k)}(T/\lambda)$ and $\mathcal{GT}^{(D,k)}(T/\mu)$ confined\\
  $\st^{(D,k)}(T,\lambda,\mu)=\log2$ &$\Leftrightarrow$& either $\mathcal{GT}^{(D,D-k)}(T/\lambda)$ or $\mathcal{GT}^{(D,k)}(T/\mu)$ deconfined\\
  $\st^{(D,k)}(T,\lambda,\mu)=0$ &$\Leftrightarrow$& both $\mathcal{GT}^{(D,D-k)}(T/\lambda)$ and $\mathcal{GT}^{(D,k)}(T/\mu)$ deconfined
 \end{tabular}
\caption{Topological entropy in $\mathcal{T}^{(D,k)}$}\label{tab:entropy}
\end{table}

\section{Discussion}\label{sec:discussion}
We have shown that the topological entropy is a good order parameter for topological order in general toric codes at finite temperature. If we denote $t^{(D,k)}_{crit} = (T/\mu)_{crit}$ the critical coupling in the $\mathcal{GT}^{(D,k)}(T/\mu)$, we can conclude that the topological entropy of $\mathcal{T}^{(D,k)}$ experiences two phase transitions at $T_{\lambda} = \lambda\,t^{(D,D-k)}_{crit}$ and $T_{\mu} = \mu\,t^{(D,k)}_{crit}$, such that 
\begin{equation}
 \st^{(D,k)}(T,\lambda,\mu) = \begin{cases}
                                2\log2\quad&\textrm{for }T<\min(T_\lambda,T_\mu)\\
				\log2\quad&\textrm{for }\min(T_\lambda,T_\mu)<T<\max(T_\lambda,T_\mu)\\
				0\quad&\textrm{for }\max(T_\lambda,T_\mu)<T.
                               \end{cases}
\end{equation}
$t^{(D,k)}_{crit}=0$ if $k=D-1$ and $t^{(D,k)}_{crit}>0$ otherwise. Thus in models with $k=1,D-1$, TE has its maximal value $S_0=2\log 2$ only at zero temperature. The well known $\mathcal{T}^{(2,1)}$ model is the only one for which $\st(T,\lambda,\mu) = 0$ for all $T>0$. Models with $1<k<D-1$ have a nontrivial low-temperature phase, in the sense that $\st^{(D,k)}(T,\lambda,\mu) = S_0$ for all $0\leq T<T_c$, $T_c = \min(T_\lambda,T_\mu)>0$. Such models only exist in $D\geq4$, the simplest example being the Kitaev's four-dimensional toric code \cite{dklp_02} ($\mathcal{T}^{(4,2)}$).

The temperature dependence of $\st$ follows the same pattern as the properties of quantum and classical memory in our systems. Indeed, if we combine Tables \ref{tab:memory} and \ref{tab:entropy}, we discover the promised connection between TE and robustness of memory in the toric codes
\begin{table}[h]
\centering
 \begin{tabular}{l c l}
  quantum memory &$\Leftrightarrow$&$\st^{(D,k)}(T,\lambda,\mu)=2\log2$\\
  classical memory &$\Leftrightarrow$&$\st^{(D,k)}(T,\lambda,\mu)=\log2$\\
  no memory &$\Leftrightarrow$&$\st^{(D,k)}(T,\lambda,\mu)=0$
 \end{tabular}
\caption{Memory and $S_{top}$ in $\mathcal{T}^{(D,k)}$}\label{tab:mement}
\end{table}

In the introduction, we mentioned the recent formulation by Hastings for topological order at $T>0$ \cite{hastings_11}. A thermal state $\rho_{eq}$ has TO if it can not be transformed arbitrarily close to a `classical' state by means of local unitaries, even if we are allowed to tensor in additional local degrees of freedom. By a classical state is meant a thermal state of a local Hamiltonian which is diagonal in a product basis. In \cite{hastings_11}, it is argued that a thermal state is topologically ordered if one can efficiently perform quantum error correction, in the sense of \cite{dklp_02}, since one can then thicken the logical operators $\bar{X}$, $\bar{Z}$ while preserving their algebraic properties. These then prevent local unitaries from conjugating the state $\rho_{eq}$ arbitrarily close to a classical state. In $\mathcal{T}^{(D,k)}$, efficient correction of defects of both kinds is possible if and only if both $\mathcal{GT}^{(D,D-k)}(T/\lambda)$ and $\mathcal{GT}^{(D,k)}(T/\mu)$ are in the confined phase. Hastings's circuit definition, the presence of stable quantum memory and the behavior of TE all agree in predicting topological order in the toric codes:

\begin{center}
\textit{A system has TO at $T>0$ if $\st(T)=\st(0)>0$.}
\end{center}

We remark again that the mere presence of non-vanishing TE is not enough to ensure TO. Indeed,  there are situations  for  classical states in the sense of  \cite{hastings_11} in which one can encode stable classical memory, for instance the toric code in 3D    $\mathcal T^{(3,1)} $, in which $\st>0$ see also \cite{classicalte}.  In order to assess TO, TE must have the full value that it has at zero temperature. In fact, up to half of that value can be of classical origin and therefore not related to quantum TO \cite{castelnovo08-3d}. 

We want to discuss the meaning of stable classical and quantum memories from the statistical mechanics point of view. The stability of classical memory corresponds to ergodicity breaking. The phase space breaks into domains in which the evolution is confined for exponentially long (in the system size) times. These domains become disconnected in the thermodynamic limit. One can encode classical information in the system by knowing in which domain the system is confined. A stable quantum memory means that there is a manifold of metastable quantum states. The breaking of ergodicity here is more dramatic, because there is an infinite number of states that are disconnected, and that have arbitrarily large overlap between each other. We find that this property is accompanied by a particular pattern of long-range entanglement, being it TE or the non triviality of entanglement (NT) criterion. We asked ourselves: Why is that? Why should the entanglement properties be related in such a strict way to the way the system approaches equilibrium?  We think that answering this question is crucial to the understanding of the notions of quantum memory, ergodicity breaking, and the statistical mechanics of topological phases.

In this paper, we have proven that  the mechanism leading to the destruction of TE is given by the confinement-deconfinement transition in the corresponding gauge theory (here the $\mathbb{Z}_{2}$ lattice gauge theory) and therefore to a critical temperature $T_c$ for the stability of $\st$. Moreover, we have proven that the same mechanism is responsible for the transition $NT\rightarrow FAC$ that describes topological order in terms of patterns of non trivial long-range entanglement. The confinement-deconfinement mechanism though, is {\em also} the one responsible for the destruction of quantum (or classical) memory \cite{dklp_02}. We can then establish that TO is a property of the way ergodicity is broken, or, in information-theoretic terms, is a property of how information encoded in the system is resilient. This is important since the entanglement criteria, or other non local order parameters \cite{diagnosing} are hardly experimentally accessible, so we need to find other properties of TO that can characterize it.

This work leaves many open questions, the most natural of all is whether these results extend to general quantum double models and the corresponding discrete gauge theories. Another generalization is to investigate this transition in general string-net models \cite{wenbook, Levin2005a}, though these do not correspond to a gauge theory. Moreover, we have seen why we obtain a sharp phase transition when we consider partitions of infinite size. For the finite system, there are exponentially suppressed branes that are able to see the topology of the partitions and therefore to give exponentially suppressed corrections to $\st$. With this in mind, we ask ourselves what is the final size scaling in order to find the critical behaviour of $\st$ near $T_c$ (scaling behaviour for the mutual information in TO was examined in \cite{scaling}). Would the critical exponents for $\st$ be related to other information-theoretic quantities, that we can perhaps measure? Moreover, in the 3D case, we see that a TE of classical origin corresponds to the existence of classical memory. Again, we wonder, what is the connection? Is there a kind of TE in all the classical phases which host a stable classical memory? A fourth question is related to the very interesting class of models whose quantum memory at finite temperature is stable for only polynomial (in the system size) lifetimes. These models are obtained by coupling with a bosonic system  \cite{hamma_09} or with other long range interactions \cite{loss} or with Hamiltonians whose excitations have fractal geometric  properties \cite{haah1, haah2, cc}. What kind of TO do these models have? What happens to their TE? And what about the Hastings' criterion for such systems? 

Along with thermal stability, the question of whether TE in itself is stable in the whole TO phase is still an open question in its generality, thought TE is known to be stable in the  $\mathbb{Z}_{2}$ lattice gauge theory \cite{hamma_08}. We know that the TO phase at $T=0$ is stable because the gap will not close for arbitrary perturbations (within a range)\cite{stability}. If TE or NT are the hallmark of TO, one needs to prove they are stable too, for the same range. We hope that the results shown in this work can also be helpful for this important question.

To finish, we want to mention that another important form of stability is dynamical stability after a quantum quench, that is a sudden dramatic change in the system Hamiltonian. It has been shown \cite{quench} that TE is stable in the quench scenario for some particular quenches in the toric code in 2D while quantum memory is not \cite{stray}. Therefore the dynamical scenario is different, and there are no results in higher dimensions. Our results in the mapping to the lattice gauge theory may prove useful to study also these scenarios.

All these questions constitute an exciting challenge for the study of novel quantum phases of matter, statistical mechanics paradigms, and quantum information theory.

\section{acknowledgments} Research at Perimeter Institute for Theoretical
Physics is supported in part by the Government of Canada through NSERC and
by the Province of Ontario through MRI.


\begin{thebibliography}{99}
\bibitem{footnote}{Here we mean the binomial distribution on the number of topological defects in $D^*_{4k}$, given the total number of topological defects in $D^*_{4k}$ and $D^*_{4k+1}$, with the constraint that both these numbers are even.}
\end{thebibliography}
\end{document}